\documentclass[12pt,a4paper]{article}
\usepackage{amsmath}
\usepackage{amssymb}
\usepackage{amsfonts}
\title{The geometrically broken object}
\author{Andrea De Martino\footnote{
Dipartimento di Fisica, Universit\`a di Roma ``La Sapienza'',
P.le A. Moro 2, 00185, Rome, Italy ; e-mail: dmart@physig.phys.uniroma1.it.}}
\date{}
\begin{document}
\maketitle
\noindent
{\small \textbf{Abstract -}}
{\small We introduce an analytically solvable model for a fragmented
object that, despite of a low degree of randomness and of the extreme
simplicity of the breaking process, displays non self-averaging
effects in its thermodynamic limit.}
\vspace{1cm}
\\
In this paper we present an exactly solvable model for a fragmented
object that, despite of the extreme simplicity of the breaking process,
displays non self-averaging effects in its thermodynamic limit.
The presence of non self-averaging quantities is a striking feature
of a variety of statistical mechanical models of disordered systems,
originating in condensed matter physics, as the low temperature spin glasses
\cite{parisi}, as well as population biology \cite{higgs}, dynamical systems'
theory \cite{derrida::rmm} and mathematics \cite{derrida::overview}.
In most cases a complete analytic study is not possible, and
the interesting probability densities are obtained numerically. The model we
propose is sensibly less random than all of the above models, and yet,
quite surprisingly, has the same statistical properties.

The question of self-averaging arises in disordered models when one
considers the statistical fluctuations of a given thermodynamic
or statistical extensive property $X$ of the model. Suppose to have
a certain system of size $N$ (under ``size'' we shall understand a
linear dimension of the system as well as the number of parts forming it,
or equally a measure of its phase space, such as the number of its possible
configurations) in which disorder is represented by some
quenched random variables. For finite $N$, $X$ is sample-dependent, in the
sense that to each sample of the system, namely
to each realization of the quenched disorder, corresponds a unique value
of $X$, and the ensemble average $\langle X\rangle$ of $X$ is obtained by
averaging over all possible realizations of the quenched disorder. The
sample-to-sample fluctuations of $X$ are described by the quantity
\begin{equation} \label{fluttuazione}
\mathcal{D}_N(X)=\frac{\langle X^2\rangle-\langle X\rangle^2}{
\langle X\rangle^2} ,
\end{equation}
which clearly depends on $N$ and is non-zero for finite $N$.
If $\mathcal{D}_N(X)\rightarrow 0$ in the thermodynamic limit
$N\rightarrow\infty$ then $X$ is said to be \emph{self-averaging} and
a sufficiently large sample is a good representative of the whole
ensemble (think of the energy of a classical ideal gas of $N$ particles).
But if $\mathcal{D}_N(X)$ tends to a finite positive value, then $X$
remains sample-dependent even in the thermodynamic limit, and
an evaluation of $X$ made on a no matter how large sample is not
significant of the value of $X$ on other samples.  If this is the case,
$X$ is said to be \emph{non self-averaging}.

What one finds in many disordered models, some of which we cited above, is that 
in the thermodynamic limit the system's phase space appears broken into
infinite basins, and that: (i) such breaking is sample dependent;
(ii) for each breaking there are finite sized basins;
(iii) the sizes of the basins are non self-averaging quantities. To see this,
one usually considers the probability $Y$ that two randomly chosen
configurations (points in phase space) belong to the same basin, 
\begin{equation} \label{Y}
Y=\sum_s W_s^2 ,
\end{equation}
where $W_s$ is the probability that a randomly chosen configuration
belongs to the $s$-th basin, and the sum is made over all basins.
To prove (ii) one has to show that $Y$'s ensemble average,
calculated over all possible ways in which the phase space may be broken, is
finite and positive, as is explained in Ref. \cite{derrida::overview}.
If one then manages to show that $Y$ lacks of self-averaging, (iii)
would soon follow. Finally, (i) is a mere consequence of (iii).
Hence one tries to demonstrate that
\begin{itemize}
\item $\langle Y\rangle>0$ (for feature (ii));
\item $\textrm{var}(Y)>0$ (for features (iii) and (i)).
\end{itemize}
The most interesting quantity to compute is the probability density
$\Pi(Y)$ of $Y$, such that $\Pi(Y)dY$ gives the probability that the
value of $Y$ for a random sample lies between $Y$ and $Y+dY$. The
fact that $Y$ has a finite variance in the thermodynamic limit tells
us that $\Pi(Y)$ remains ``broad'', and this is confirmed by computer
simulations, but an analytic derivation of $\Pi(Y)$ is not possible
in most models. Along with $\Pi(Y)$, other interesting quantities are
the probability densities of the weights of the largest, second largest,
\ldots pieces in each sample, that are expected to remain ``broad'' as well,
and the average number of pieces of a given size for a sample.

The ``broken phase space'' feature explains why fragmentation models
may attract some interest. One assumes that an object of size $1$ is
broken into infinite pieces by means of a certain process, which depends on
a number of quenched random variables. A realization of the breaking
process is given by a specific sampling of the random variables and
the result of each breaking is an infinite set of pieces of sizes $W_s$
($s=1,2,\ldots$). Averages are then calculated over all possible ways of
breaking the object.
Of course, not all breaking processes lead to non self-averaging
effects. Suppose for example that the object is broken uniformly into $M$
basins of size $W_s\sim\frac{1}{M}$ each; then $Y\sim\frac{1}{M}$, and both
$W_s$ and $Y$ would go to zero in the $M\rightarrow\infty$ limit.
In some sense, the fact that a uniformly broken object is self-averaging
suggests that a non-uniformly broken object, such
as the randomly broken object \cite{derrida::rbo}, may be regarded as the
simplest complex system. We shall now see that a much simpler, though
similar, breaking process than that of the randomly broken object leads
to comparable non self-averaging effects, with the advantage of
being analytically solvable.

We consider an object of size $1$ and
suppose to break it into infinite pieces by means of the following process:
given a real number $p\in[0,1]$, we tear the object in two pieces of sizes
$1-p$ and $p$ respectively; then we do the same thing with the piece of size
$p$, obtaining two pieces of sizes $(1-p)p$ and $p^2$ respectively, plus
the one of size $1-p$ obtained at the first breaking step.  If we repeat
the same procedure with the pieces of sizes $p^2$, $p^3$, \ldots that
are obtained at the second, third, \ldots breaking steps, keeping the
pieces of sizes $(1-p)p$, $(1-p)p^2$, \ldots obtained at the same steps
we finally have a set of pieces of sizes
\begin{eqnarray}
W_1 & = & 1-p\nonumber\\
W_2 & = & (1-p)p\nonumber\\
\ldots &   &\\
W_s & = & (1-p)p^{s-1}\nonumber\\
\ldots, &   &\nonumber 
\end{eqnarray}
where $W_s$ denotes the size of the piece kept at the $s$-th step.
Clearly, $\sum_s W_s=\sum_{s=1}^{\infty}(1-p)p^{s-1}=1$.
We call such process a \emph{geometrical breaking}, since the sizes of the
resulting pieces form a geometric sequence. Now suppose that $p$ is
a random variable with probability density $\rho(p)$. In this case we
can imagine that a breaking sample is produced by choosing a random
value of $p$ from $\rho(p)$, so that averaging over all samples means
simply averaging over all possible values of $p$.

As usual, let $Y=\sum_s W_s^2$. $Y$'s value for a single sample is given by
\begin{equation} \label{Ydip}
Y=\sum_s (1-p)^2 p^{2(s-1)}=\frac{(1-p)^2}{1-p^2}=\frac{1-p}{1+p},
\end{equation}
and the ensemble average of $Y$ is simply
\begin{equation}
\langle Y\rangle=\int_0^1\frac{1-p}{1+p}\rho(p)dp .
\end{equation}
When $\rho$ is uniform we easily obtain the value
\begin{equation} \label{Yprimo}
\langle Y\rangle=\log 4 -1 \simeq 0.386\ldots .
\end{equation}

The probability density $\Pi(Y)$ over the samples may be calculated from
the relation $\Pi(Y)dY=\rho(p)dp$
expressing the conservation of probability. From formula (\ref{Ydip}) we
get $p=(1+Y)^{-1}(1-Y)$ and thus
\begin{equation}
\Pi(Y)=\bigg|\frac{dp}{dY}\bigg|\rho\bigg(\frac{1-Y}{1+Y}\bigg) ,
\end{equation}
namely
\begin{equation}
\Pi(Y)=\frac{2}{(1+Y)^2}\rho\bigg(\frac{1-Y}{1+Y}\bigg).
\end{equation}
$Y$'s ensemble average is given by $\langle Y\rangle=\int_0^1 Y\Pi(Y) dY$
and one can verify that for a uniform $\rho$ the value (\ref{Yprimo})
is recovered. It suffices to use the fact that
\begin{equation} 
\int\frac{2Y}{(1+Y)^{2}}dY=2\bigg(\log|1+Y|+\frac{1}{1+Y}\bigg)+\textrm{constant}.
\end{equation}
The second moment $\langle Y^2\rangle=\int_0^1Y^2\Pi(Y)dY$
may also be calculated in a straightforward manner for a uniform $\rho$.
We can make use of the relation
\begin{equation} 
\int\frac{2Y^2}{(1+Y)^{2}}dY=2\bigg(-\frac{Y^2}{1+Y}+2\Big(1+Y-\log|1+Y|\Big)\bigg)+\textrm{constant}
\end{equation}
to obtain $\langle Y^2\rangle=3-\log 16\simeq 
.227\ldots\neq\langle Y\rangle^2$. The variance of $Y$ is finally given by
\begin{equation} \label{Ysecondo}
\textrm{var}(Y)=\langle Y^2\rangle-\langle Y\rangle^2\simeq 0.078\ldots.
\end{equation}
The fact that $Y$ has a non zero variance in the thermodynamic limit, that for
this model is represented by the infinite pieces in which the object is
broken, proves that $Y$, and consequently the sizes $W_s$, lack of
self-averaging for a geometrically broken object. We remind that all
numerical values were obtained under the assumption of a uniform density
$\rho(p)$.

Let us now turn to the other interesting quantities. Since for each sample one
may find a piece of maximum size, then a second maximum sized piece and so
on, one might be interested in the densities of the probability that the
$n$-th largest piece in a sample has size between $W$ and $W+dW$. Let us
denote such densities by $P_n(W)$, so that $P_1(W)dW$ represents the
probability that the largest piece in a sample has size between $W$ and
$W+dW$, $P_2(W)dW$ represents the probability that the second largest piece
in a sample has size between $W$ and $W+dW$ and so on. For a geometrically
broken object one can easily understand that, for each sample,
\begin{equation}
W_1>W_2>\cdots
\end{equation}
since, for $p\in[0,1]$, it is always $1-p>(1-p)p>(1-p)p^2>\cdots$ .
Hence, the largest size is $W=1-p$, so that, from the relation
$P_1(W)dW=\rho(p)dp$ and from the fact that, for the largest piece,
we have $p=1-W$, follows
\begin{equation} \label{pi1diw}
P_1(W)=\bigg|\frac{dp}{dW}\bigg|\rho(1-W)=\rho(1-W) .
\end{equation}
For the second largest piece one has $W=(1-p)p$. According to a standard
method of probability theory we can write
\begin{equation}
P_2(W)=\frac{\rho(p_+)}{|W'(p_+)|}+\frac{\rho(p_-)}{|W'(p_-)|}
\end{equation}
where $p_+$ and $p_-$ are the roots of the second order polynomial
$(1-p)p-W=0$, and $W'(p)=\frac{dW}{dp}=1-2p$. We have
\begin{equation}
p_{\pm}=\frac{1}{2}\Big(1\pm\sqrt{1-4W}\Big) ,
\end{equation}
which leads to
\begin{equation} \label{pi2diw}
P_{2}(W)=\frac{1}{\sqrt{1-4W}}\Bigg[\rho\bigg(\frac{1}{2}\Big(1+\sqrt{1-4W}
\Big)\bigg)+\rho\bigg(\frac{1}{2}\Big(1-\sqrt{1-4W}\Big)\bigg)\Bigg].
\end{equation}

In principle, one can calculate the densities $P_n(W)$ for every $n$ by
solving the polynomial $(1-p)p^{n-1}-W=0$ and applying the technique used
above in its generalization given by the formula
\begin{equation}
P_n(W)=\sum_{i=1}^{n}\frac{\rho(p^{(i)})}{|W'(p^{(i)})|} ,
\end{equation}
where $p^{(i)}$ is the $i$-th root of the polynomial.

The average largest, second largest, \ldots sizes are obtained by
$\langle W_n\rangle=\int_0^1W P_n(W) dW$ which may be as well written
\begin{equation}
\langle W_n\rangle=\int_0^1 (1-p)p^{n-1}\rho(p)dp .
\end{equation}
For a uniform $\rho$ this leads to
\begin{equation}
\langle W_n\rangle=\frac{1}{n}-\frac{1}{n+1}=\frac{1}{n(n+1)}
\end{equation}
($n=1$ corresponds to the average largest size, $n=2$ to the average second
largest size, and so on), so that on the average the largest piece is
three times as big as the second largest, six times bigger than the
third largest, \ldots, fortyfive times as big as the $9$-th largest, and
so on\footnote{
This type of distribution of the average weights is known as
Zipf's law (M. Gell-Mann, Zipf's law and related mysteries,
lecture held at the SFI Complex Systems Winter School, Tucson,
January 12-24, 1992 (unpublished)).}.

We can also calculate the average number of pieces of size between $W$ and
$W+dW$, denoted by $f(W)$. We know that the $n$-th largest piece has size
$W_n=(1-p)p^{n-1}$. This size is maximum for
$p=p^{(\textrm{max})}=\frac{n-1}{n}$,
in which case it is given by
\begin{equation}
W_n^{(\textrm{max})}=W_n(p^{(\textrm{max})})=\frac{(n-1)^{n-1}}{n^n} .
\end{equation}
This means that each $P_n(W)$ is defined just for $W<W_n^{(\textrm{max})}$
(in fact, $P_2(W)$ is defined for $W<\frac{1}{4}=W_2^{(\textrm{max})}$). If
in a certain sample we have a piece of size $\frac{1}{4}<W<1$ it must be the
largest piece in the sample, since the second largest piece can have a
maximum size of $\frac{1}{4}$. Hence, for $\frac{1}{4}<W<1$ one has
\begin{equation}
f(W)=P_1(W)=\rho(1-W) ,
\end{equation}
by means of (\ref{pi1diw}). If we have a piece of size $\frac{4}{27}<W
<\frac{1}{4}$ it can be either the largest one or the second largest one,
since the third largest one has a maximum size of $\frac{4}{27}$. Hence, for
$\frac{4}{27}<W<\frac{1}{4}$ one has
\begin{eqnarray} 
f(W) & = & P_1(W)+P_2(W)\nonumber\\
& = & \rho(1-W)+\frac{1}{\sqrt{1-4W}}\Bigg[\rho\bigg(\frac{1}{2}\Big(1+
\sqrt{1-4W}\Big)\bigg)+\\
&   & +\rho\bigg(\frac{1}{2}\Big(1-\sqrt{1-4W}\Big)\bigg)\Bigg]\nonumber ,
\end{eqnarray}
by means of (\ref{pi1diw}) and (\ref{pi2diw}).
In the same way, if we have a piece of size
$\frac{n^n}{(n+1)^{n+1}}<W<\frac{(n-1)^{n-1}}{n^n}$
(for any positive integer $n$) this can be any
of the first $n$ largest pieces, since the $(n+1)$-th largest piece can have a
maximum size of $\frac{n^n}{(n+1)^{n+1}}$. Hence for $\frac{n^n}{(n+1)^{n+1}}
<W<\frac{(n-1)^{n-1}}{n^n}$ one has
\begin{equation}
f(W)=P_1(W)+\cdots +P_n(W) .
\end{equation}
This shows how one can reconstruct the function $f(W)$ in all intervals of the
form $\frac{n^n}{(n+1)^{n+1}}<W<\frac{(n-1)^{n-1}}{n^n}$, and hence for all
$W\in]0,1]$.

To summarize, we have shown that this geometrically broken object
has the same statistical properties as other, well known disordered
models, and displays lack of self-averaging. Furthermore, the
simplicity of its definition makes it possible to solve it exactly, and we
have derived all the most interesting quantities, such as the probability
densities $\Pi(Y)$, $P_1(W)$ and $P_2(W)$, and the average number of
pieces of size $W$, $f(W)$. We believe that this model is interesting
because the breaking process involves just one random variable, whereas
in the randomly broken object the number of random variables entering
the breaking process is infinite. In this sense, the geometrically
broken object is less disordered than other disordered models, and
the fact that comparable non self-averaging effects are present is
quite surprising.

We have applied the results obtained in this paper to
the study of a stochastic evolutionary model \cite{adm},
in which the phase space breaks like a geometrically broken object,
to show for example that the model lacks of self-averaging in the
thermodynamic limit.

\end{document}